\documentclass[review]{elsarticle}
\usepackage{graphicx}
 \usepackage[usenames]{color}
\usepackage{lineno,hyperref}
\usepackage{lineno}
\usepackage{geometry}
\journal{JINST}

\begin{document}


\begin{frontmatter}
\title{{\bf Design and performance of a large-area scintillator-based chamber for the MID subsystem of ALICE 3}}

\author[if_unam]{Ruben~Alfaro~Molina}
\author[tecnm]{Juan~Carlos~Cabanillas~Noris}
\author[chicago]{Edmundo~Garc\'ia~Solis} 
\author[icn_unam]{Laura~Helena~Gonz\'alez~Trueba}
\author[if_unam]{Varlen~Grabski}
\author[cinvestav]{Gerardo~Herrera~Corral}
\author[icn_unam]{Jes\'us~Eduardo~Mu\~noz~M\'endez}
\author[uas]{Ildefonso~Le\'on~Monz\'on}
\author[icn_unam]{Antonio~Ortiz}
\author[icn_unam]{Antonio~Paz}
\author[icn_unam]{Ian~P\'erez~Garc\'ia}
\author[icn_unam]{Ricardo~Rodr\'iguez Pineda}
\author[prague]{Solangel~Rojas~Torres}
\author[fcfm_buap]{Guillermo~Tejeda~Mu\~noz}
\author[icn_unam]{Paola~Vargas~Torres}
\author[icn_unam]{Victor~V\'azquez~Campos}
\author[fcfm_buap]{Yael~Antonio~V\'asquez~Beltran}
\address[if_unam]{Instituto de F\'isica, Universidad Nacional Aut\'onoma de M\'exico, Mexico}
\address[tecnm]{Instituto Tecnol\'ogico de Culiac\'an, Tecnol\'ogico Nacional de M\'exico, Mexico}
\address[chicago]{Chicago State University, Chicago, USA}
\address[icn_unam]{Instituto de Ciencias Nucleares, Universidad Nacional Aut\'onoma de M\'exico, Mexico}
\address[cinvestav]{Centro de Investigación y Estudios Avanzados, Mexico City, Mexico}
\address[uas]{Universidad Aut\'onoma de Sinaloa, Mexico}
\address[prague]{Czech Technical University in Prague, Prague, Czech Republic}
\address[fcfm_buap]{Facultad de Ciencias F\'isico Matem\'aticas, Benem\'erita Universidad Aut\'onoma de Puebla, Mexico}

%
%
%




%
%


\cortext[mycorrespondingauthor]{antonio.ortiz@nucleares.unam.mx}


\begin{abstract}

This paper reports on the design and construction of a chamber for the muon identifier detector (MID) of the ALICE~3 upgrade project. The chamber consists of two sensitive layers separated by a 1\,cm air gap. Each layer holds 24  scintillator bars ($1\times4\times100$\,cm$^3$) manufactured by FNAL-NICADD. The bars are equipped with Kuraray wavelength shifting fibers and the readout is provided by a silicon photomultiplier from Hamamatsu. The bars in the second layer are orthogonal to the bars in the first layer, thus providing an overlapping cell size of 4$\times$4\,cm$^{2}$. The bar assembly as well as the design of the mechanical structure is described. The design of the chamber is close to that considered in the ALICE~3 letter of intent. The chamber was tested at the CERN T10 beamline using 3\,GeV/$c$ pion-enriched and muon beams. The chamber was placed behind an iron absorber, with different absorber lengths considered in the test. The muon identification is performed using a Machine Learning algorithm, which was trained and tested using muon (signal) and pion (background) data (50\% of the available statistics). The trained ML algorithm was applied to muon data, yielding a muon efficiency above 99\% for the OR condition (detection in either layer 1 or 2). The implementation in the pion-beam data gives the fake-muon efficiency as a function of the absorber length that is well described by an exponential function with a slope parameter of 18.79 cm. The next steps towards finalizing the optimization are outlined.   

\end{abstract}

\begin{keyword}
\texttt{Upgrade,} \texttt{Scintillators,} \texttt{SiPM,} \texttt{Machine Learning,} \texttt{ALICE experiment} 
\end{keyword}

\end{frontmatter}

\section{Introduction
\label{sec:Introduction}}
The ALICE apparatus has been successful in studying  the properties of the strongly-interacting quark--gluon plasma (sQGP), the deconfined state of strongly-interacting matter~\cite{ALICE:2022wpn,Bala:2016hlf}. However, some fundamental questions on the sQGP properties and other aspects of the strong interaction will remain open after the LHC Runs 3 and 4 (2022-2033)~\cite{alice3loi}. To address these questions and fully exploit the potential of the heavy-ion collisions at the LHC during Run 5 (2036-2041), a completely new apparatus named ALICE\,3 is proposed~\cite{Dainese:2925455}. Among all the subsystems for ALICE~3, the muon identifier detector (MID) is included. It consists of 180 chambers installed on the surface of a standard magnetic iron absorber covering around 4 hadronic interaction lengths (70\,cm). The main goal is the reconstruction of $J/\psi$ at rest in the dimuon-decay channel, therefore, MID is optimized to identify muons with momentum within 1.5-5\,GeV/$c$. This feature will make ALICE~3 unique because other experiments at the LHC can only identify muons above 6\,GeV/$c$. For MID, one of the technologies that is under consideration is based on scintillator bars. Simulations using the PYTHIA~8 event generator and GEANT~4 for the detector simulation and propagation of particles through the materials were performed. The results suggest that a time resolution of a few ns is relevant to suppress slow particles that might interact with the detector. Position resolution of a few cm is also sufficient since it is limited by scattering in the absorber. Moreover, based on simulations, only two layers made of scintillator bars will be enough to reconstruct muons with an efficiency of 94\% and fake-muon efficiency below 3\%. The detector performance has been experimentally confirmed at CERN T10 using a small chamber prototype~\cite{MunozMendez:2025ttk}.     

This paper reports on the large chamber design (including the mechanical design), experimental setup at CERN T10, and the performance of the analysis using Machine Learning. The paper is organized as follows: in section~\ref{sec:DetectorPrototype} the construction of the prototype is discussed. Section~\ref{sec:ExperimentalSetup} presents the experimental setup. Section~\ref{sec:Methodology} discusses the data analysis using Machine Learning, and finally, section~\ref{sec:Results} discusses the results.

\section{Prototype construction
\label{sec:DetectorPrototype}}

The prototype consists of 48 scintillator paddles, with 24 mounted on an aluminum plate and the other half on a second plate. This allows for a gap between the two layers of 1\,cm, and also a bar spacing in each layer of 0.1\,cm. The scintillator was manufactured by FNAL-NICADD~\cite{Grachov:2004jg}. According to previous studies, this low-cost material is enough for the MID requirements~\cite{MunozMendez:2025ttk,Alfaro:2024sxc}.   The dimensions of the bars are 100$\times$4$\times$1\,cm$^3$, they incorporate a lengthwise groove in the middle where a wavelength-shifting fiber (WLS) is coupled with optical cement (Eljen EJ-500), and a silicon photomultiplier (SiPM) attached to one side of the fiber. The paddle assembly procedure, as well as the details of the mechanical structure, are outlined below.

\subsection{Paddle assembly
\label{sec:layout}}

\begin{figure}[h!]
    \centering
    \includegraphics[width=0.5\linewidth]{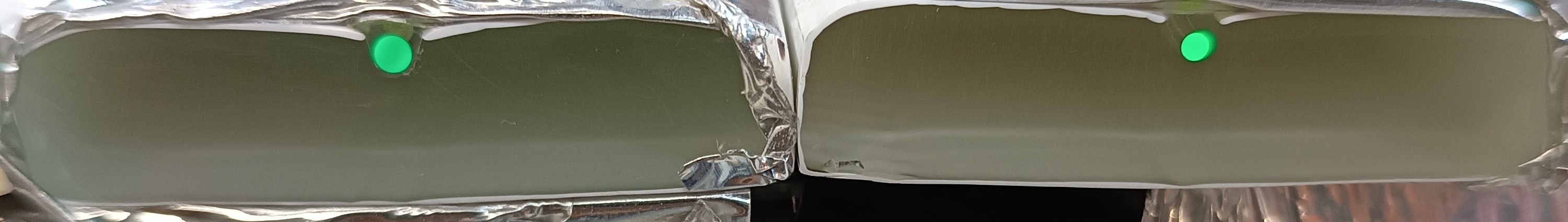}
    \caption{The polished surfaces for the 2.0\,mm (left) and 1.5\,mm (right) WLS fibers already attached to the scintillator.}
    \label{fig:fib2mm}
\end{figure}

The process starts with the cleaning of the groove using isopropyl alcohol, the quality is verified through a visual inspection with a microscope. Optical cement is used to glue the WLS fiber in the groove. After 24 hours, the optical cement is already dry, which allows polishing the bar- and fiber-end surfaces. Two different WLS-fiber diameters (Kuraray Y-11) were used: 1.5\,mm and 2\,mm. The light yield was found to increase by 40\% from the 1.5\,mm to the 2.0\,mm case. Figure~\ref{fig:fib2mm} shows the scintillator surfaces after the polishing treatment.

\begin{figure}[h!]
    \centering
    \includegraphics[width=0.5\linewidth]{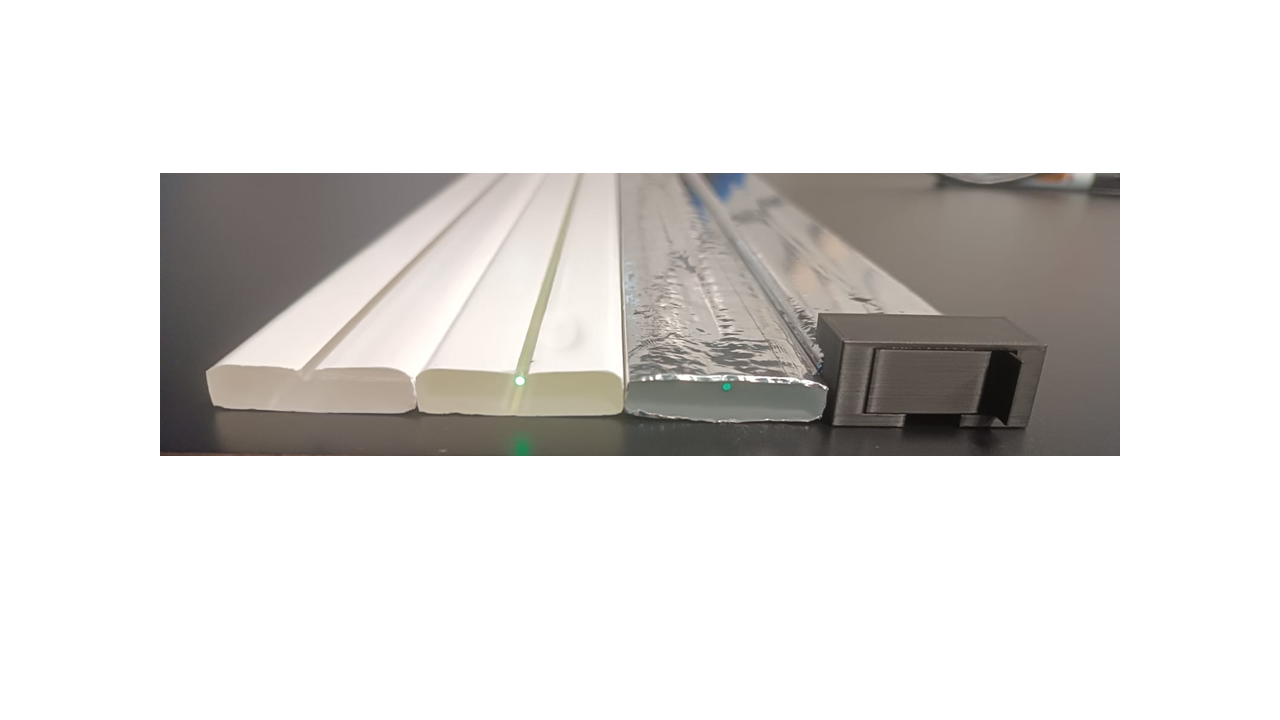}
    \includegraphics[width=0.45\linewidth]{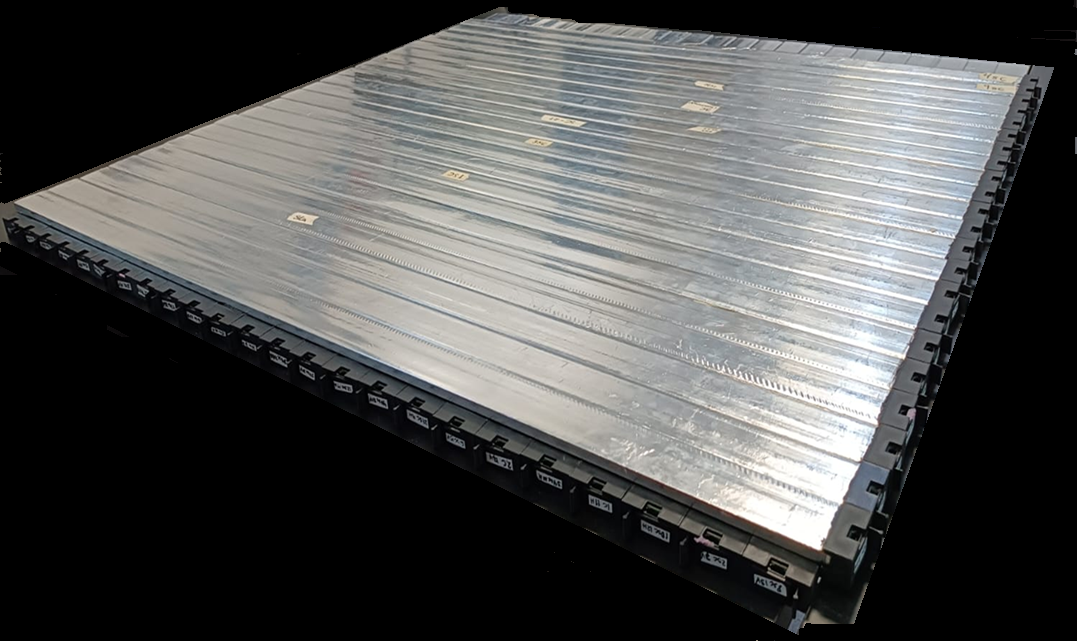}    
    \caption{(Left) Assembly steps for the bars from left to right: from a bar without a fiber to the bar already wrapped with the SiPM holder. (Right) 48 assembled bars with their snap-fit holders for SiPMs, ready to be introduced to the mechanical structure.}
    \label{fig:fasesdeconstruccion}
\end{figure}

The paddle is wrapped using aluminum tape 3M-3311 (50\,$\mu$m thick).  To collect the light transmitted through the fiber, a Hamamatsu SiPM (model 14160-3050HS, active area $3\times 3$\,mm$^2$) is used. A printed circuit board (PCB) is assembled to the connector and the SiPM. To fix the PCB with the SiPM to the bar, a snap-fit holder was used. The holder is manufactured in a 3D printer and avoids screws in the mechanical structure.  Figure~\ref{fig:fasesdeconstruccion} shows the steps for bar assembly, from the bar without fiber to the bar with the holder. The holders were fixed to the bar using optical cement mixed with carbon powder. A 1:1 ratio of optical cement with carbon was used. The same figure also displays the 48 bars, already assembled and ready to be introduced into the mechanical structure; each bar with aluminum tape has a final width of $\approx41$\,mm.

\subsection{Mechanical structure
\label{sec:mechanics}}

\vspace{0.5cm}

\begin{figure}
    \centering
    \includegraphics[width=0.5\linewidth]{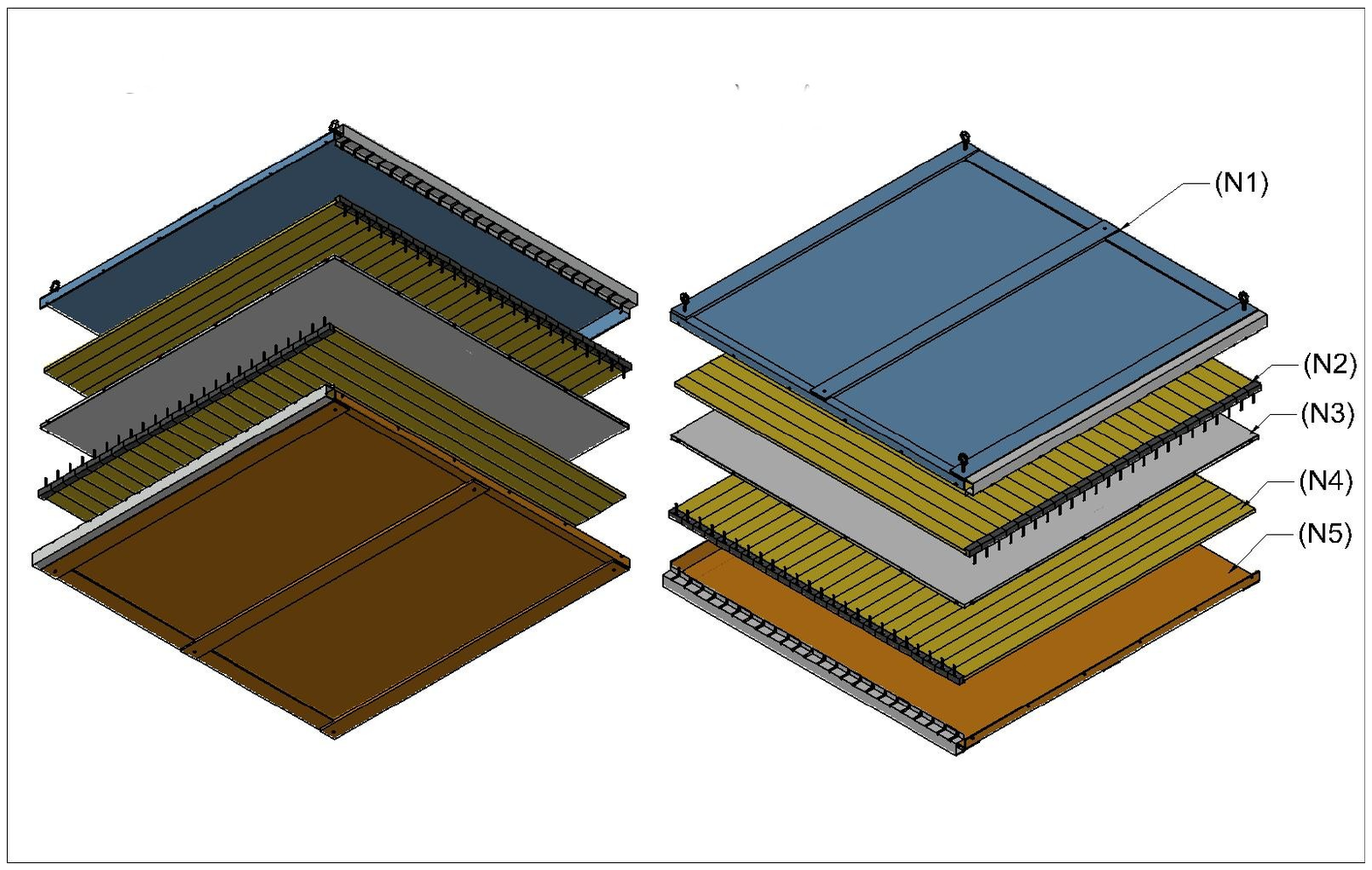}
    \caption{Mechanical structure for the $1\times1$m$^2$ prototype, where N2 and N4 are the scintillator layers separated by N3 by 1 cm.}
    \label{fig:estrucmecanica}
\end{figure}

\vspace{0.5cm}
The mechanical framework of the chamber is constructed using sills and angle supports that hold the aluminum plates; these plates, in turn, support the scintillator bars. Figure \ref{fig:estrucmecanica} illustrates the design of this structure. In this configuration, N1 corresponds to the top aluminum cover, N2 to the first scintillator layer, N3 to the intermediate aluminum plate separating the two scintillator layers, N4 to the second scintillator layer, and N5 to the bottom aluminum cover \cite{Caja_reporte}.

\begin{figure}
    \centering
    \includegraphics[width=0.5\linewidth]{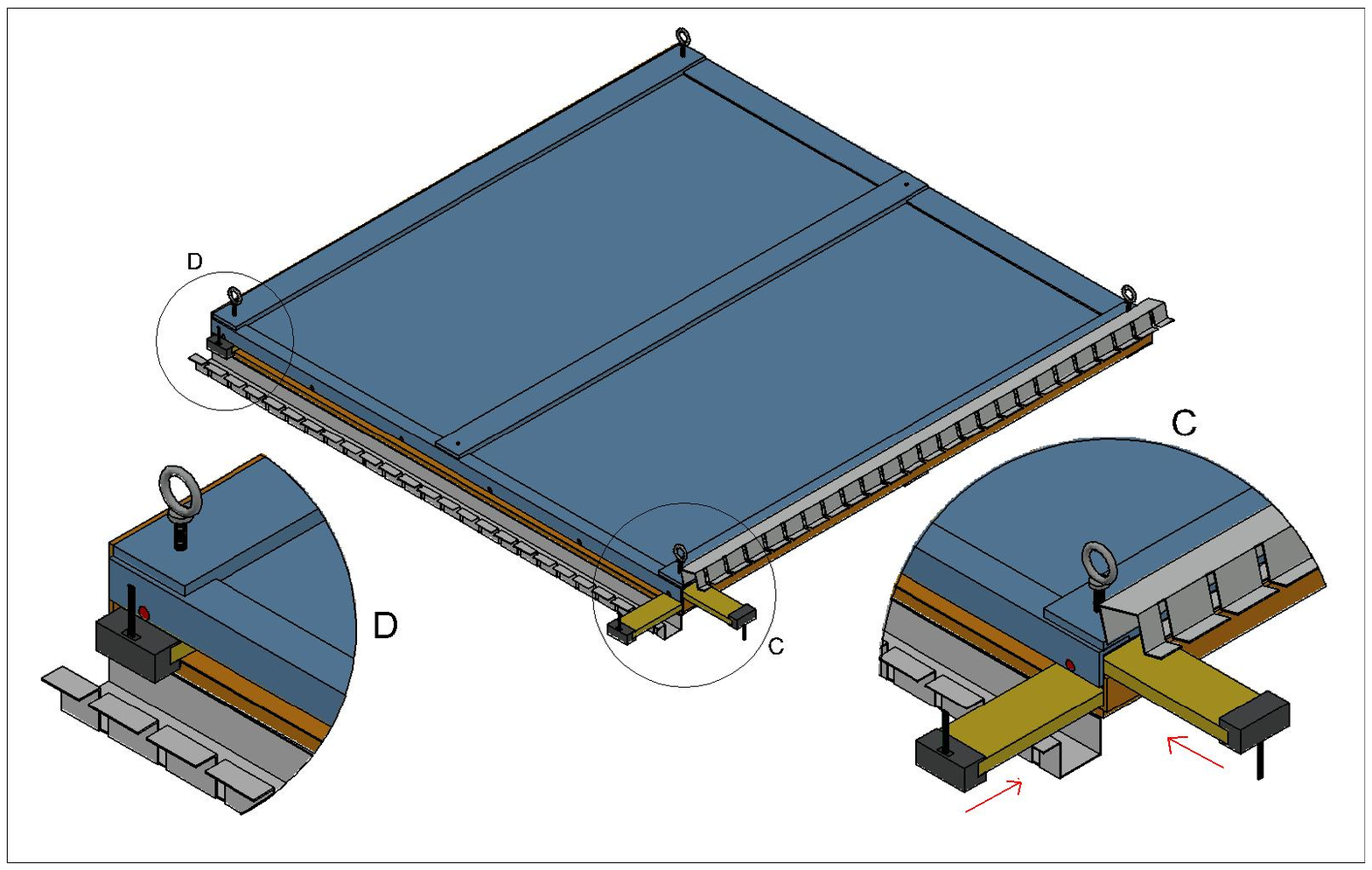}
    \caption{Complete design for the mechanical structure, showing how each bar is arranged in its place, with the SiPM protection holders.}
    \label{fig:estructura_prototipo}
\end{figure}

\begin{figure}
    \centering
    \includegraphics[width=0.45\linewidth]{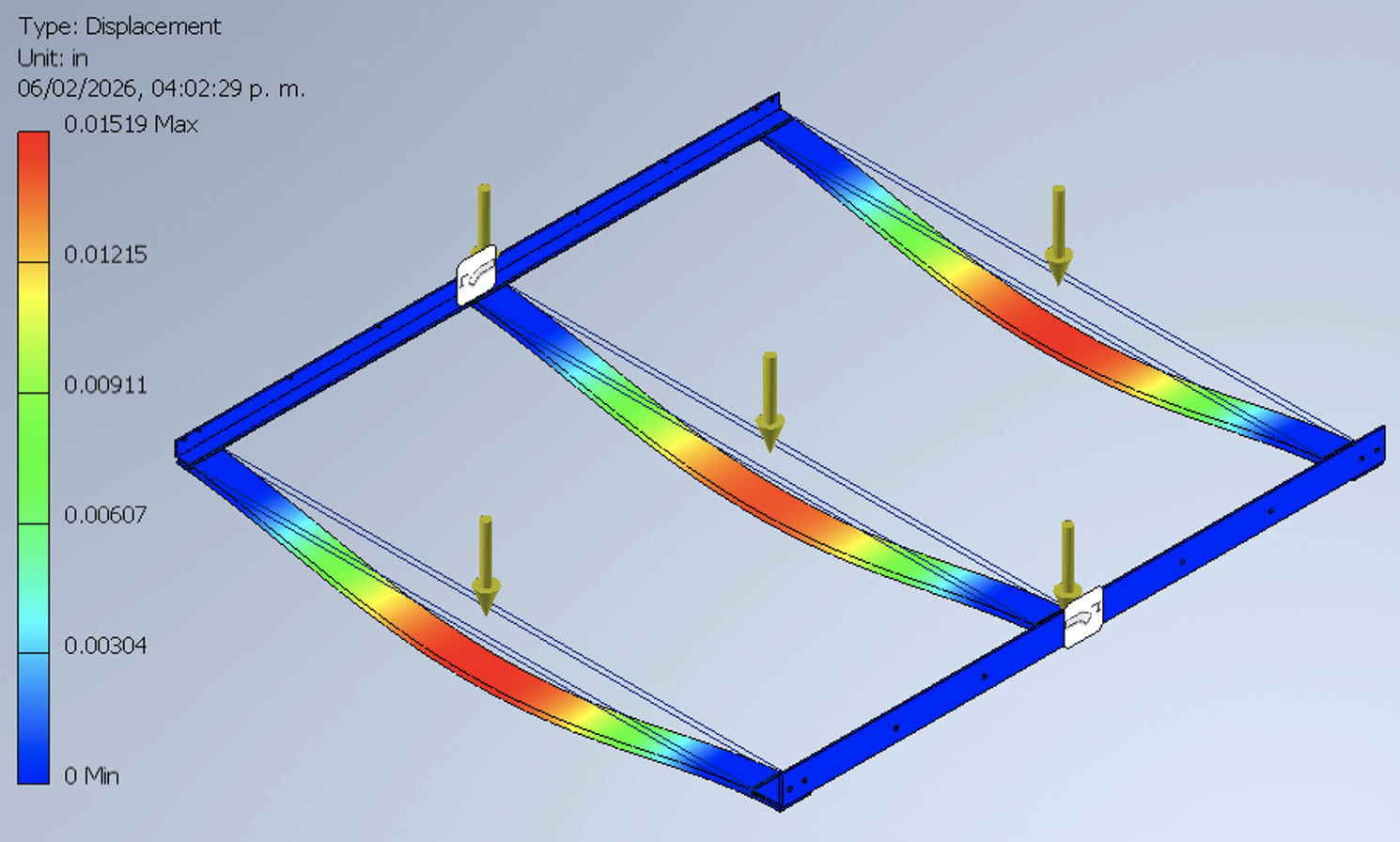}
    \includegraphics[width=0.45\linewidth]{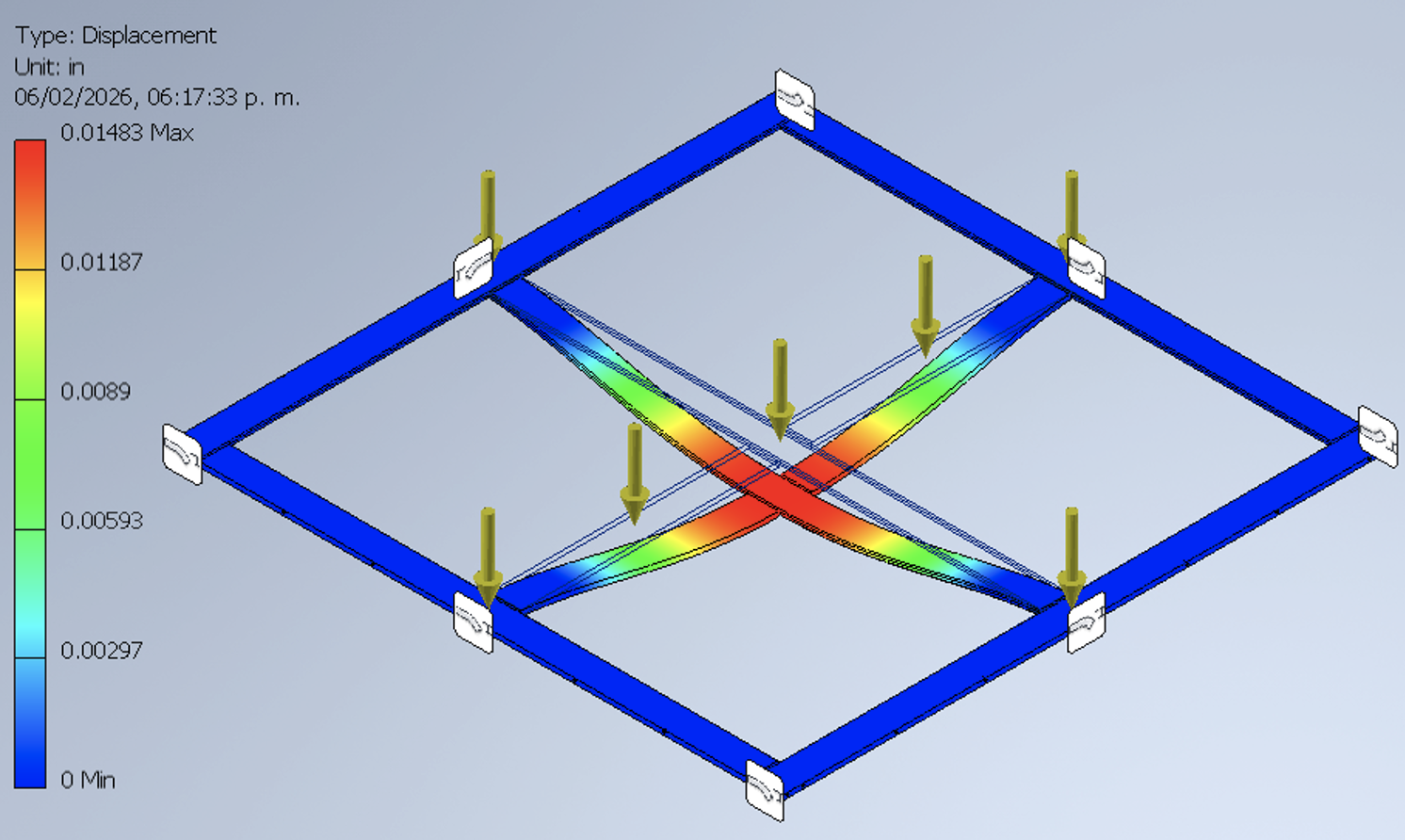}    
    \caption{Simulation of the mechanical stress for the top and bottom aluminum cover (left), and for the intermediate aluminum plate (right).}
    \label{fig:stress}
\end{figure}

Figure \ref{fig:estructura_prototipo} shows the complete assembled structure. Its final outer dimensions are $1033.2\times 1033.2 \times 56.4$ mm$^3$. The sills supporting the chamber are approximately $2 × 1/8$ in$^2$, and two of the angle profiles share these dimensions. The other two angles, located on the sides near the scintillator-bar holders, have dimensions of $1 × 1/8$ in$^2$. The aluminum layers are made from 20-gauge sheets with a surface area of $1\times1$ m$^2$. To ensure that the structure has the mechanical strength to support the bars some simulations were made. The results are shown in Fig.~\ref{fig:stress}, the simulated mechanical load (10kg) on the top cover (N1 and N5) for scintillators and 2.73\,kg for 1\,m$^{2}$ of 19-gauge sheet metal. For the analysis, two sides were fixed, and the total load was uniformly distributed. The maximum deflection of 0.386\,mm was found at the center. For the middle surface (N3) the maximum deflection is 0.377\,mm,  assuming a uniform mechanical load of 10\,kg (scintillators) plus 5.46\,kg (2\,m$^{2}$ of 19-gauge sheet metal).

\section{Experimental setup
\label{sec:ExperimentalSetup}}

\begin{figure}[h!]
    \centering
    \includegraphics[width=0.9\linewidth]{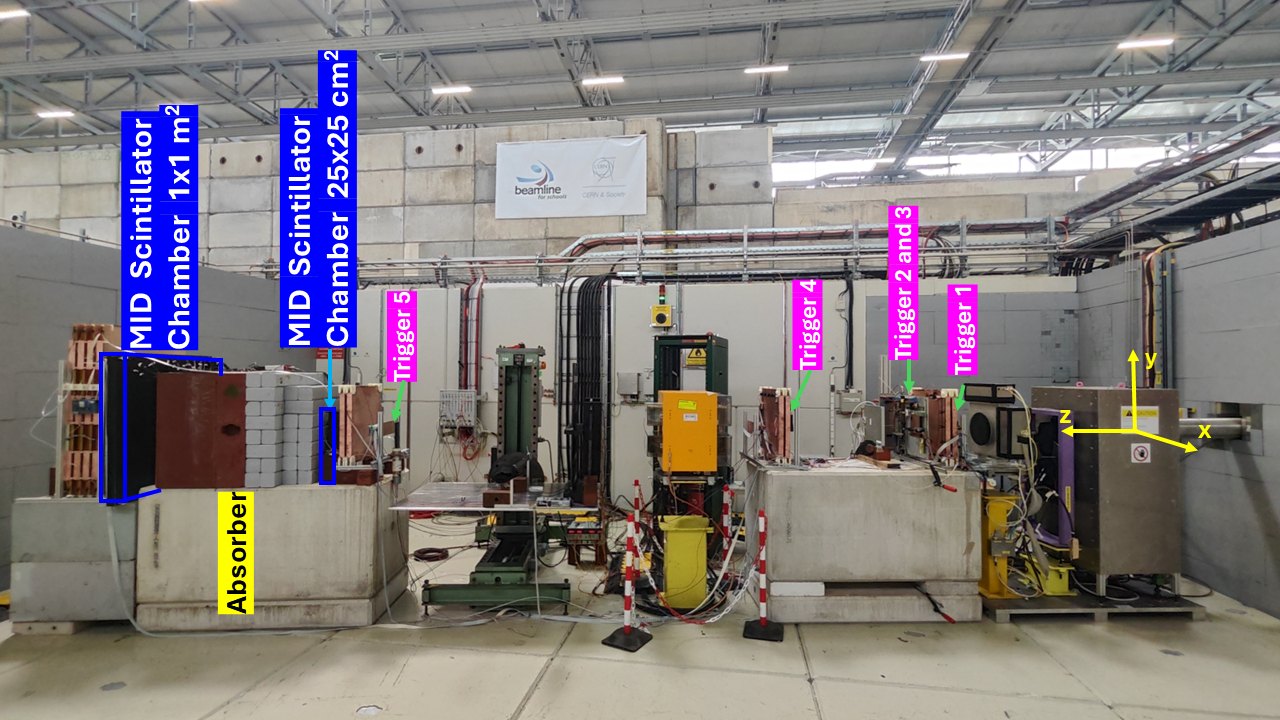}
    \caption{Experimental setup. The trigger scintillators are labeled as Trigger 1, ..., Trigger 5. The beam travels from right to left along the $z$ coordinate, which is perpendicular to the array of detectors
tested at T10.}
    \label{fig:setup}
\end{figure}

The MID chamber was tested with 3\,GeV/$c$ $\pi^{-}$-enriched and $\mu^{+}$ beams. Figure~\ref{fig:setup} shows the setup installed in the beam area. The setup consisted of five scintillator paddles to provide a clean trigger, an iron absorber of $80\times80$\,cm$^{2}$ transverse area of variable length (from 46 to 86 cm) followed by the MID chamber. The coordinate system is defined as follows: the beam particles travel along the $z$ axis, the horizontal axis corresponds to the $x$ axis, whereas the vertical axis corresponds to the $y$ axis.

The trigger signal was provided by a coincidence of the signals from five scintillator paddles (hereinafter called trigger scintillators), which were located in front of the absorber. They are indicated as Trigger 1-5; triggers 1 and 5 have an active transverse area of $7\times10$ cm$^2$, triggers 2 and 3 have an area of $5\times1$ cm$^2$, and trigger 4 has an area of $2.8\times2.8$ cm$^2$. The position of the paddles with respect to the beam is indicated in Fig.~\ref{fig:setup}. The trigger signals were processed with a NIM leading-edge discriminator (CAEN N840). The output signals were further processed with a coincidence module (LeCroy 622) to produce the 5-fold coincidence that defined the trigger signal. In order to clean the data from the incoming beam and have a reference time-of-arrival (ToA$_{\rm ref}$) for the measurements with the large prototype, a small scintillator chamber was included in the readout. Only events with signals in the center of the small scintillator chamber were processed. To remove electron contamination reported in Refs.~\cite{vanDijk:2025ggb,MunozMendez:2025ttk} a high-pressure threshold Cherenkov counter (XCET) in the T10 beam line was used, providing a cleaner beam than in previous studies.  This contamination amounts to $0.290 \pm 0.002$ for the negative-charged pion-enriched beam.  A pressure scan was carried out in order to identify the threshold pressure at which only pions could be isolated, and this was found to be 2.4\,bar. The signal from the XCET was connected to the veto input of the CAEN N405 logic unit to produce a 2-fold coincidence signal with the previously mentioned 5-fold coincidence to exclude the electron background.

\section{Data analysis
\label{sec:Methodology}}

In the 2024 test-beam campaign~\cite{MunozMendez:2025ttk} the methodology to achieve a high-muon efficiency was outlined. The analysis strategy relied on the use of a Machine Learning algorithm, specifically, the boosted decision trees (BDT) implemented in TMVA (Toolkit for Multivariate Data Analysis~\cite{Hocker:1019880}). The training and testing phase used GEANT~4 simulations. The input variables were time-of-arrival (ToA), position in layer 1, position in layer 2, and time-over-threshold (ToT). However, for the 2025 MID test beam campaign, all triggered events were recorded because the readout electronics (DT5202 module) enabled collection of both charge and time, and a small chamber was placed in front of the absorber. Therefore, for the present analysis, the number of hits per event ($N_{\rm hit}$) is available, and for each hit the following information can be used: charge, ToA-ToA$_{\rm ref}$ and positions (quantified in terms of the fired channel). Since clean muon and pion beams were achieved using the high-pressure threshold, the Cherenkov detector data could be used for the training and testing phase.  

\begin{figure}[h!]
    \centering
    \includegraphics[width=0.8\linewidth]{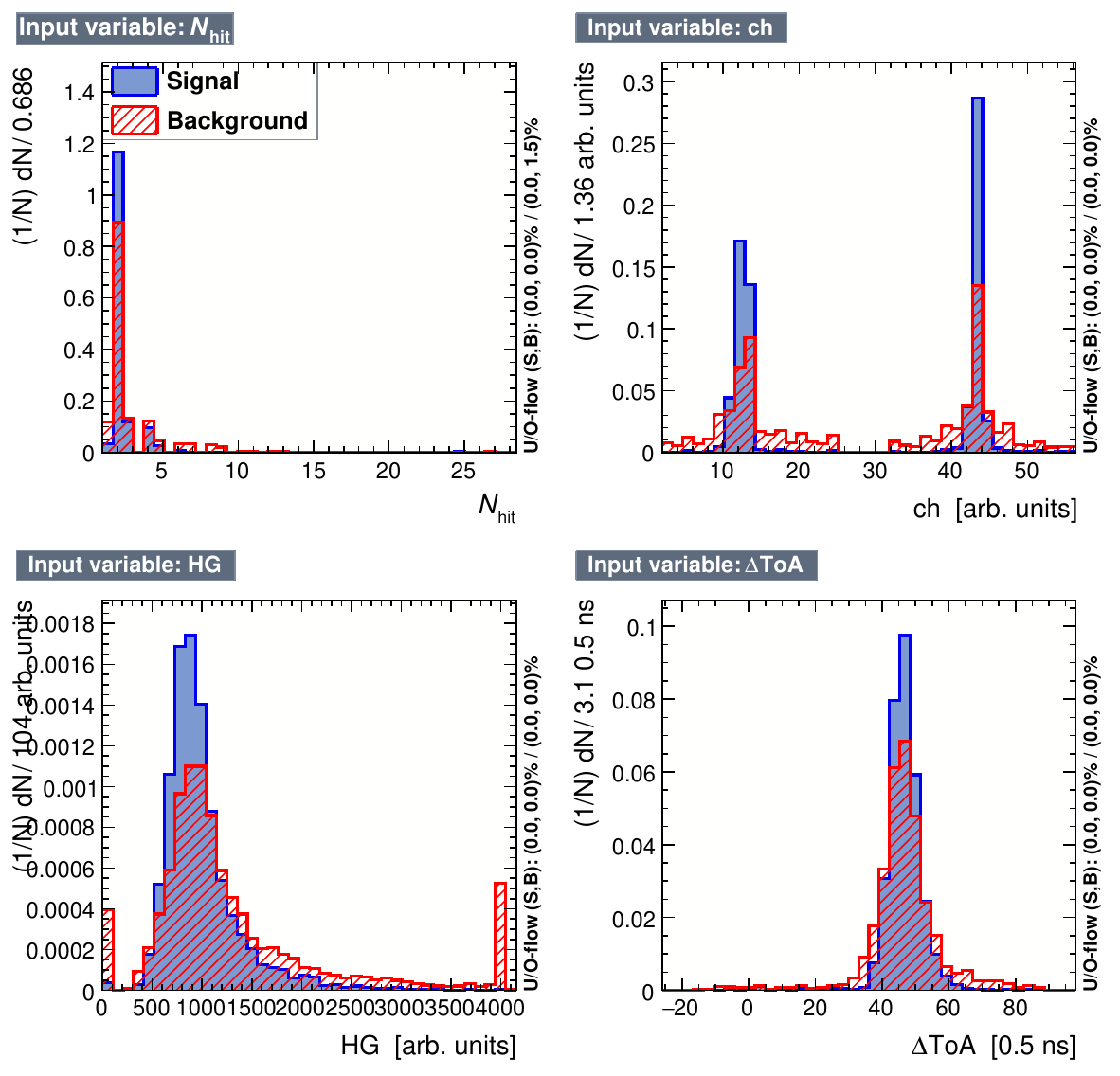}
    \caption{Input variables used for training: hit multiplicity (upper-left plot), fired channel (upper-right plot), charge (bottom-left plot) and $\Delta$ToA=ToA-ToA$_{\rm ref}$ (bottom-right plot). The signal (3\,GeV/$c$ muons) and background (3\,GeV/$c$ pions) are displayed in blue and red shaded areas, respectively. Results for an absorber of 76\,cm effective thickness are shown.}
    \label{fig1:analysis}
\end{figure}

The training/testing phase used 50\% of the available statistics from both the muon (signal) and the hadron (background) beams. The remaining data were used to measure the hadron suppression factor and its corresponding muon efficiency. The input variables for training are shown in Fig.~\ref{fig1:analysis} for hits selected within a time window of 50\,ns around the nominal expected time, and for the absorber thickness of 76\,cm. While the fraction of events with a single hit is 18.0\% for the pion beam,  4.3\% is observed for the muon beam. This quantifies the fraction of particles hitting only a single layer. However, excluding the first bin, the average hit multiplicity is 3.0 (standard deviation 1.7) for the pion beam and 2.7 (standard deviation 0.52) for the muon beam.   Moreover, both the position and charge distributions are wider for the pion beam than for the muon beam.  These behaviours are fully consistent with the presence of hadron showers in the pion beam.

\section{Results
\label{sec:Results}}

\begin{figure}[h!]
    \centering
    \includegraphics[width=0.45\linewidth]{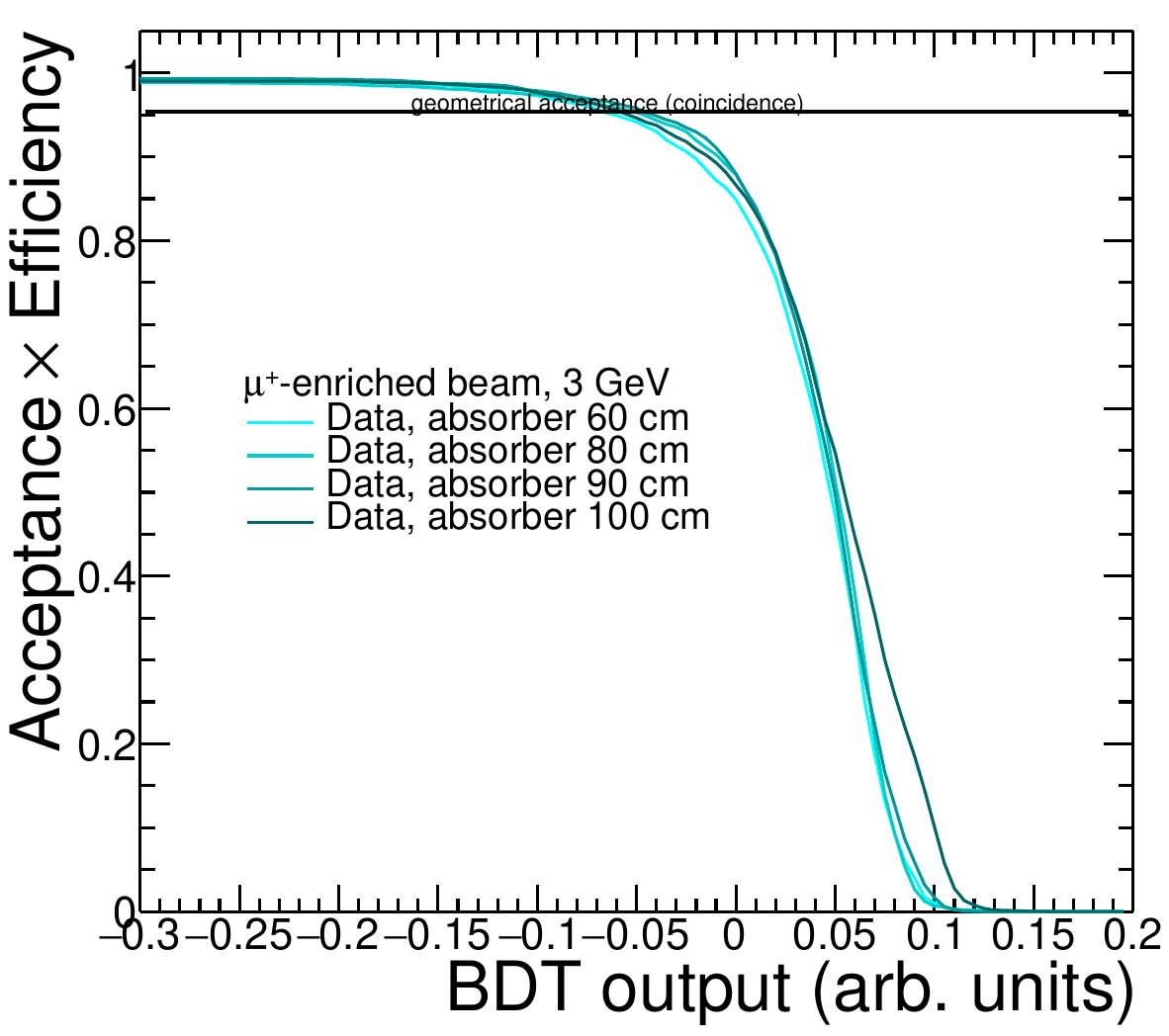}
    \caption{Geometrical acceptance times efficiency as a function of the BDT output. Results for all the absorber lengths are shown.}
    \label{fig2:analysis}
\end{figure}

The geometrical acceptance times efficiency as a function of the BDT output for different absorber lengths is shown in Fig.~\ref{fig2:analysis}. The result is presented for the muon beam. For low BDT-output values ($<-0.25$), the quantity is around 99\%, which is close to the OR condition case in which at least one layer is fired. However, one has to make a compromise between a good muon efficiency and a competitive hadron suppression factor.  

\begin{figure}[h!]
    \centering
    \includegraphics[width=0.45\linewidth]{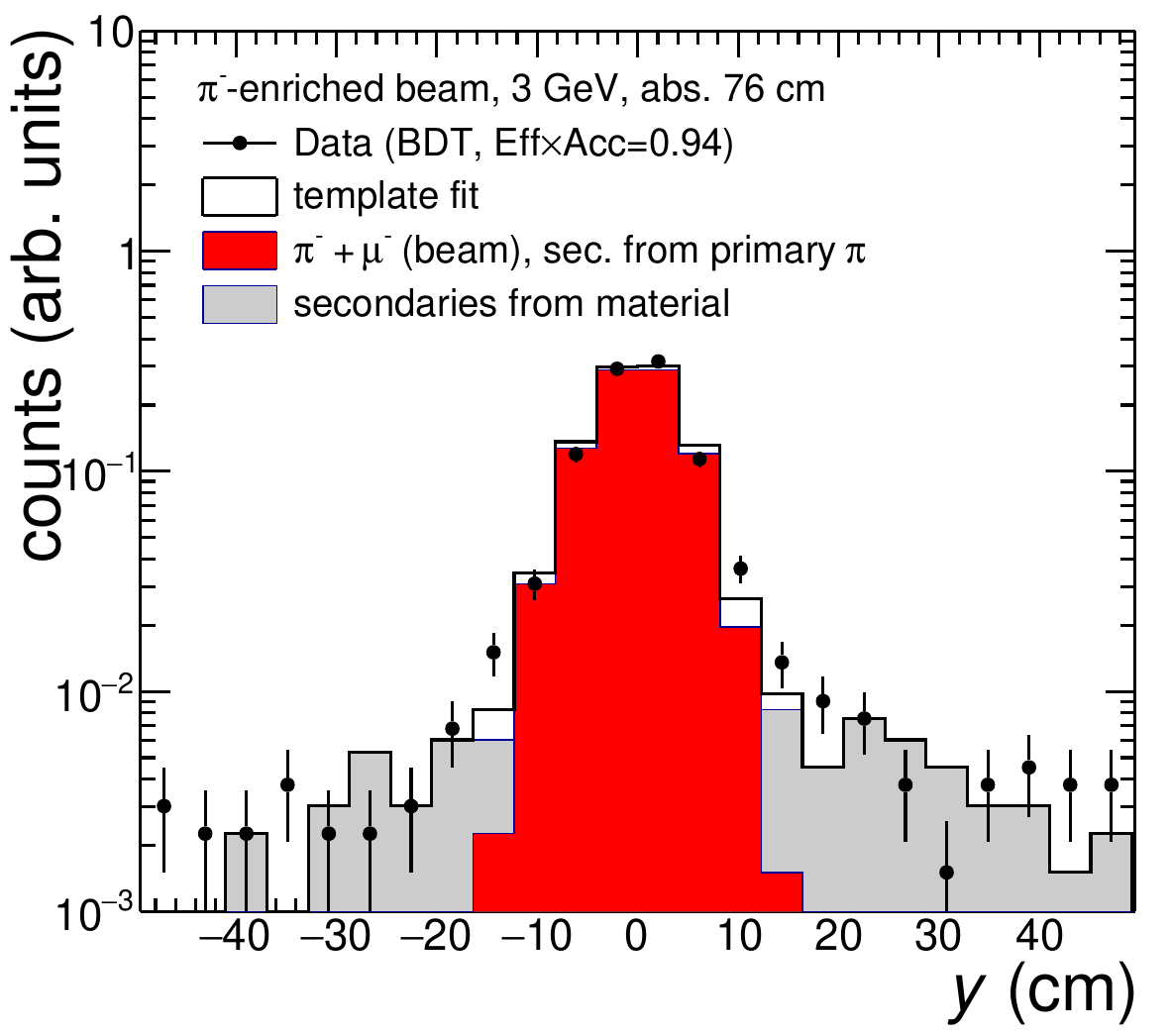}
    \includegraphics[width=0.45\linewidth]{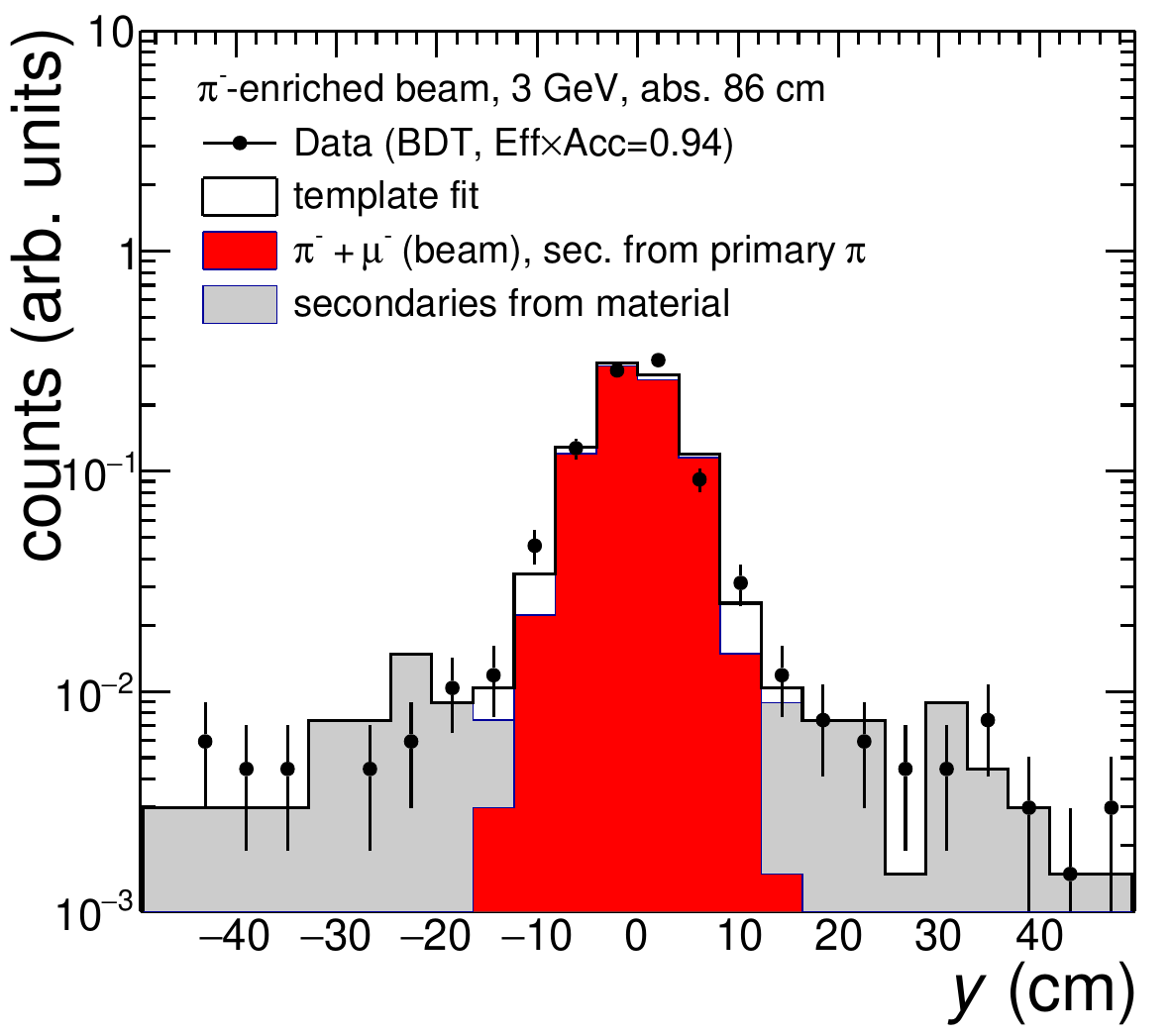}    
    \caption{Hit-position distribution measured for the 3\,GeV/$c$ pion-enriched beam. Hits are selected considering a BDT-output threshold at which the muon acceptance$\times$efficiency is 94\%. A two-Gaussian function is fitted to data. Results for an absorber thickness of 76 (left) and 86\,cm (right) absorber length are displayed.}
    \label{fig3:analysis}
\end{figure}

Figure~\ref{fig3:analysis} shows the $y$ position (in layer 1) distribution of hits selected from the pion-enriched sample after the implementation of the trained BDT. The hit selection is done using the BDT-output threshold at which the muon acceptance $\times$ efficiency is 94\%. The pion-enriched-beam data are well described by a convolution of two Gaussian distributions, the first with a narrow width corresponding to primary-particle candidates and a second (with a wider width) associated with secondary particles. This parametrization is based on the shape of the position distributions expected from GEANT~4 simulations~\cite{MunozMendez:2025ttk}. Using such a parametrization, the fraction of primary candidates from a template-fit method is around 85\% for an absorber thickness of 76\,cm. This survival particle yield normalized to the number of triggers and is referred to as ``fake-muon efficiency''. 

\begin{figure}[h!]
    \centering
    \includegraphics[width=0.45\linewidth]{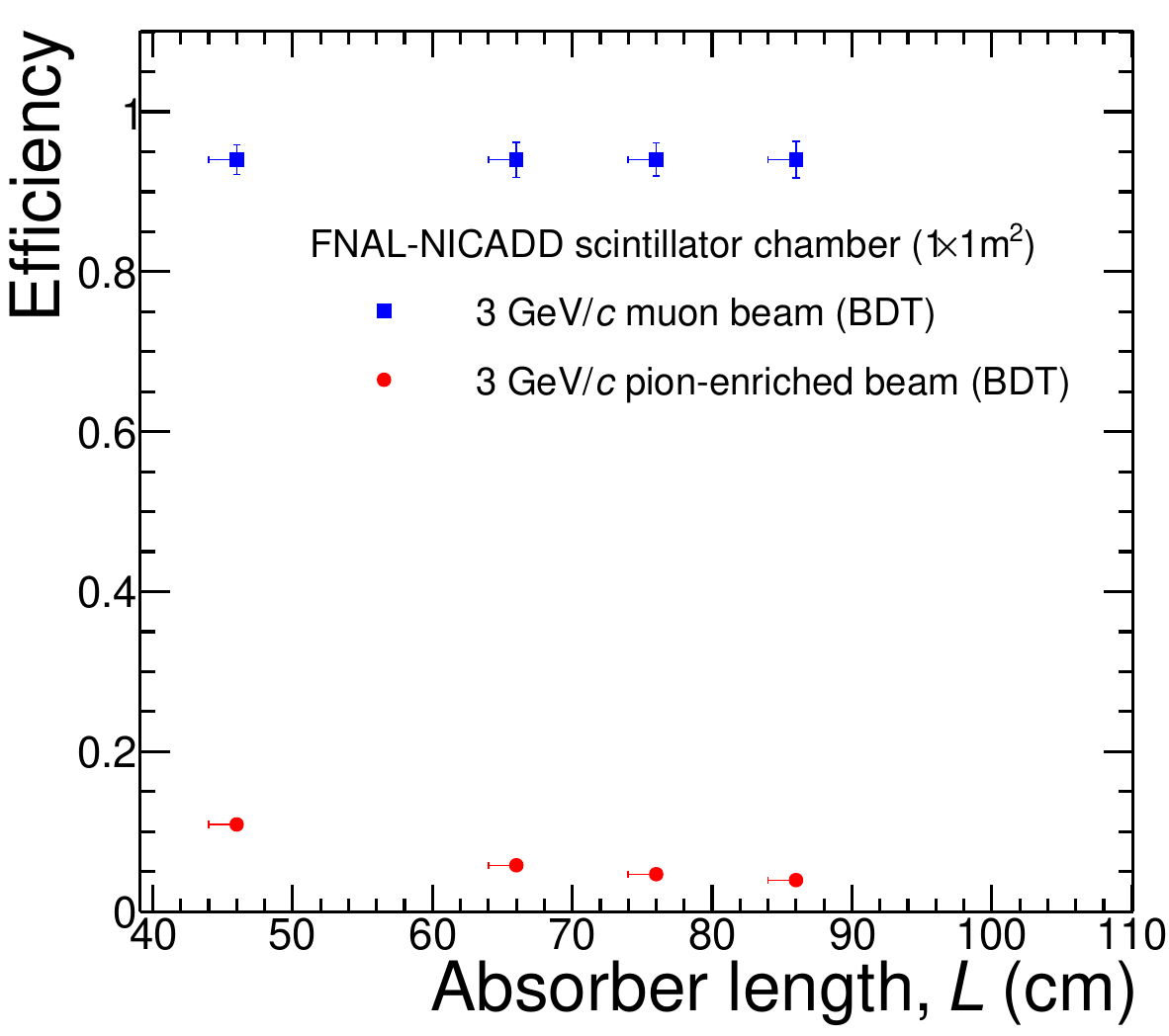}
    \includegraphics[width=0.45\linewidth]{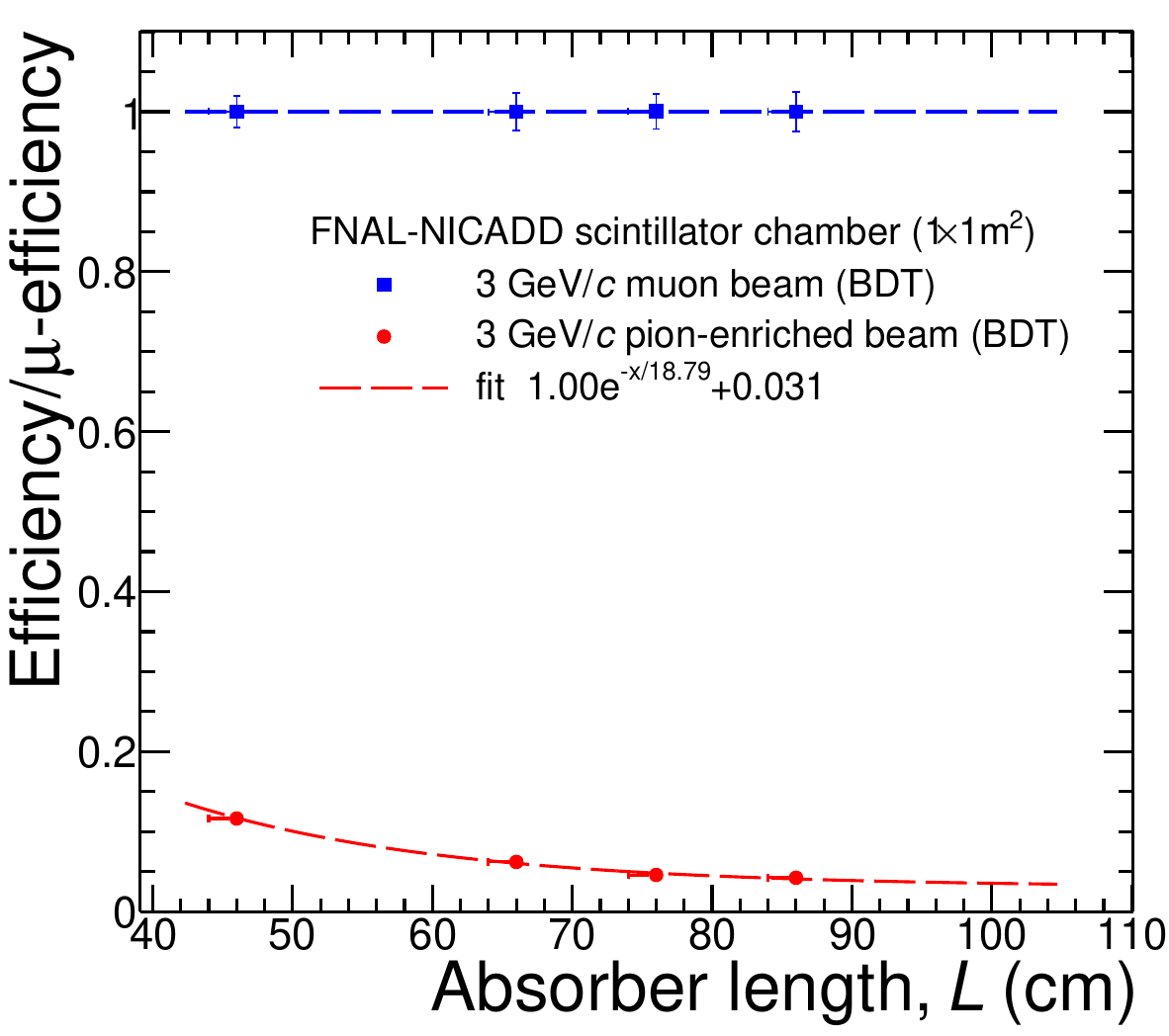}    
    \caption{Muon-candidate efficiency as a function of absorber length (left). Results for the pion-enriched (red round markers) and muon (blue squares) beam are displayed. An exponential convoluted with a constant function is fitted to data (dashed line). The fraction of identified muon candidates is also shown (right).}
    \label{fig4:analysis}
\end{figure}

The muon and the fake-muon efficiency as a function of absorber length is shown in Fig.~\ref{fig4:analysis}. By construction, the muon acceptance $\times$ efficiency as a function of absorber length is around 94\%. However, the fake muon efficiency stays below 10\% for an absorber thickness above 50\,cm. The right-hand side plot of Fig.~\ref{fig4:analysis} illustrates the fraction of observed particles, which are essentially the efficiencies described above normalized to the muon efficiency. By construction, 100\% of muons were recovered, but what is more relevant is the fraction of observed fake-muons. The pion data were well modeled by an exponential function plus a constant term ($e^{-x/\lambda_{\rm I}}+b$, with $\lambda_{\rm I}=18.79$\,cm and $b$=0.031). The constant contribution of 3.1\% can be due to a remaining muon contamination in the pion-enriched beam. For example, the pion-decay in flight probability in 8\,m (air) is 4.6\,\%. This contribution is reduced due to the trigger condition and therefore, a small percentage of muon content is expected.  


\begin{figure}[h!]
    \centering
    \includegraphics[width=0.45\linewidth]{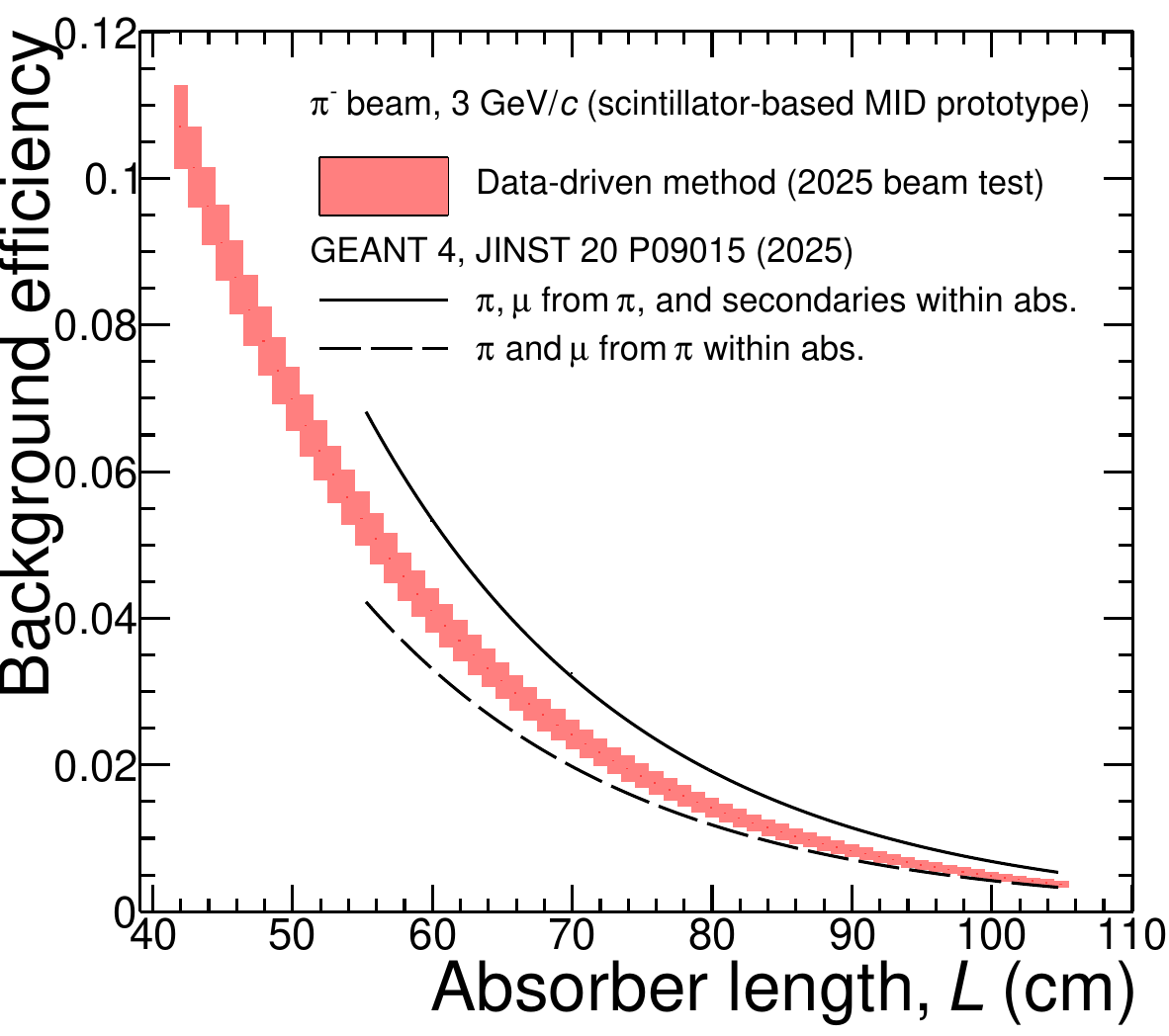}   
    \caption{Data-driven background efficiency as a function of absorber length. The 2025-test-beam data are compared with the GEANT~4-based background efficiency extracted from the analysis of 2024-test-beam data.}
    \label{fig5:analysis}
\end{figure}

A data-driven method is adopted to extract the muon contamination, essentially, a constant factor of 3.1\% is subtracted. Figure~\ref{fig5:analysis} shows the background efficiency as a function of the absorber length. The  band represents an uncertainty given by the uncertainty in the absorber thickness. This uncertainty is due to a small hole (about 2\,cm)  that is placed at the center of the absorber. An exponential plus a constant function was fitted to data assuming the lowest and highest absorber length values. For the reference absorber length, 70\,cm, the background efficiency amounts to $2.4\pm0.1$\%. The same figure also shows a comparison with the background efficiency reported in a previous paper considering a smaller MID prototype with the same type of scintillator and WLS fiber~\cite{MunozMendez:2025ttk}. In such a reference, the efficiency relied on GEANT~4 simulations. Using pure muon (signal) and pion (background) beams, the BDT were trained. The algorithm was further applied both to data and GEANT~4 simulations. The background efficiency includes contributions from primary pions, muons from pion decays and other secondaries (from material). The figure illustrates a comparison with two cases: a) background including all contributions, and b) background including only primary pions and muons from pion decays. The data-driven background efficiency using the big prototype stays below the case (a) and above the case (b). This is because in the analysis of 2025 beam-test data, a template-fit method was implemented in order to further suppress the contribution from other secondaries. 

\section{Summary and outlook \label{sec:Conclusions}}

The construction of a large-detection-area muon chamber prototype ($1\times1$\,m$^{2}$) is discussed. The prototype consists of 48 scintillator bars equipped with WLS fibers and readout with SiPMs. The detector has two layers, each one hosting 24 scintillator bars. The bar spacing is 0.1\,mm and the gap between the layers is 1\,cm.  The chamber was placed behind a cast-iron absorber of different thickness. The specifications of the detector fulfill those described both in the ALICE~3 letter of intent as well as in the ALICE~3 scoping document. 

The test is performed at the CERN T10 facility, a threshold Cherenkov counter was used to suppress the electron contamination in both the 3\,GeV/$c$ pion- and muon-enriched beams. The results indicate that operating the detector at a muon efficiency of 94\% would be sufficient to give a fake-muon efficiency of 2.4\% for a 70\,cm-thick absorber. In the next months, new MID prototypes consisting of longer scintillator bars (1.5\,m) will be tested, as well as scintillator from different vendors. 


\appendix


\section*{Acknowledgments}
The authors acknowledge the technical support from Luciano D\'iaz Gonz\'alez, Jes\'us Eduardo Murrieta Le\'on, Sa\'ul Aguilar Salazar, and Jaime Everardo P\'erez Rodr\'iguez. The MID-chamber construction/test, and data processing were supported through funding from PAPIIT-UNAM (Grant IG100524) and PAPIME-UNAM (Grant PE100124). S. R. acknowledges the Czech Agency: support by the Ministry of Education, Youth and Sports of the Czech Republic project LM2023040. 


\bibliography{ref}

@article{MunozMendez:2025ttk,
    author = "Mu{\~n}oz M{\'e}ndez, Jes{\'u}s Eduardo and others",
    title = "{ML-based muon identification using a FNAL-NICADD scintillator chamber for the MID subsystem of ALICE 3}",
    eprint = "2507.02817",
    archivePrefix = "arXiv",
    primaryClass = "physics.ins-det",
    doi = "10.1088/1748-0221/20/09/P09015",
    journal = "JINST",
    volume = "20",
    number = "09",
    pages = "P09015",
    year = "2025"
}

@article{alice3loi,
    collaboration = "ALICE",
    title = "{Letter of intent for ALICE 3: A next-generation heavy-ion experiment at the LHC, CERN-LHCC-2022-009, LHCC-I-038}",
    eprint = "2211.02491",
    archivePrefix = "arXiv",
    primaryClass = "physics.ins-det",
    reportNumber = "CERN-LHCC-2022-009, LHCC-I-038",
    month = "11", 
    year = "2022"
}

@techreport{Caja_reporte,
    author = "Pérez Rodríguez, J.E.",
    title = "{Mecánica del MID (Identificador de Muones), Proyecto ALICE3 CERN}",
    institution = "Instituto de Física, UNAM" ,
    year = "2025"
}

@article{vanDijk:2025ggb,
    author = "van Dijk, Maarten and others",
    title = "{Particle production and identification for the T10 secondary beamline of the CERN East Area}",
    eprint = "2507.02567",
    archivePrefix = "arXiv",
    primaryClass = "hep-ex",
    doi = "10.1016/j.nimb.2025.165907",
    journal = "Nucl. Instrum. Meth. B",
    volume = "569",
    pages = "165907",
    year = "2025"
}

@techreport{Dainese:2925455,
      author        = "Dainese, A and Di Mauro, A",
      title         = "{Scoping document for the ALICE 3 detector}",
      institution   = "CERN",
      reportNumber  = "CERN-LHCC-2025-002, LHCC-G-185",
      address       = "Geneva",
      year          = "2025",
      url           = "https://cds.cern.ch/record/2925455",
}

@article{Alfaro:2024sxc,
    author = "Alfaro, Ruben and others",
    title = "{Characterisation of plastic scintillator paddles and lightweight MWPCs for the MID subsystem of ALICE 3}",
    eprint = "2401.04630",
    archivePrefix = "arXiv",
    primaryClass = "physics.ins-det",
    doi = "10.1088/1748-0221/19/04/T04006",
    journal = "JINST",
    volume = "19",
    number = "04",
    pages = "T04006",
    year = "2024"
}

@inproceedings{Grachov:2004jg,
    author = "Grachov, Oleg A. and others",
    title = "{Study of new FNAL-NICADD extruded scintillator as active media of large EMCal of ALICE at LHC}",
    booktitle = "{11th International Conference on Calorimetry in High-Energy Physics (Calor 2004)}",
    eprint = "physics/0405028",
    archivePrefix = "arXiv",
    reportNumber = "FERMILAB-CONF-04-046",
    doi = "10.1142/9789812701978_0007",
    pages = "49--54",
    month = "5",
    year = "2004"
}

@article{ALICE:2022wpn,
    author = "Acharya, Shreyasi and others",
    collaboration = "ALICE",
    title = "{The ALICE experiment: a journey through QCD}",
    eprint = "2211.04384",
    archivePrefix = "arXiv",
    primaryClass = "nucl-ex",
    reportNumber = "CERN-EP-2022-227",
    doi = "10.1140/epjc/s10052-024-12935-y",
    journal = "Eur. Phys. J. C",
    volume = "84",
    number = "8",
    pages = "813",
    year = "2024"
}

@article{Bala:2016hlf,
    author = "Bala, Renu and Bautista, Irais and Bielcikova, Jana and Ortiz, Antonio",
    title = "{Heavy-ion physics at the LHC: Review of Run I results}",
    eprint = "1605.03939",
    archivePrefix = "arXiv",
    primaryClass = "hep-ex",
    doi = "10.1142/S0218301316420064",
    journal = "Int. J. Mod. Phys. E",
    volume = "25",
    number = "07",
    pages = "1642006",
    year = "2016"
}

@techreport{Hocker:1019880,
      author        = "Hocker, Andreas and others",
      title         = "{TMVA - Toolkit for Multivariate Data Analysis with ROOT:
                       Users guide. TMVA - Toolkit for Multivariate Data
                       Analysis}",
      institution   = "CERN",
      reportNumber  = "physics/0703039, CERN-OPEN-2007-007",
      address       = "Geneva",
      year          = "2007",
      url           = "https://cds.cern.ch/record/1019880",
      note          = "TMVA-v4 Users Guide: 135 pages, 19 figures, numerous code
                       examples
  and references",
}

\end{document}